\documentclass[aps,prd,reprint,superscriptaddress,nofootinbib,longbibliography]{revtex4-2}
\usepackage[T1]{fontenc}
\usepackage[utf8]{inputenc}
\usepackage{amsmath,amssymb,bm}
\usepackage{graphicx}
\usepackage{xcolor}
\usepackage{hyperref}
\hypersetup{colorlinks=true,linkcolor=blue,citecolor=blue,urlcolor=blue}

\newcommand{\dd}{\mathrm{d}}
\newcommand{\GeV}{\mathrm{GeV}}
\newcommand{\tr}{\mathrm{Tr}}
\newcommand{\Order}{\mathcal{O}}

\begin{document}

\title{Pion and kaon D terms from holographic QCD and coupled-channel
dispersion relations}

\author{Zhibo Liu}
\email{Contact author: zhibo.liu.hep@gmail.com}
\affiliation{Kobayashi-Maskawa Institute for the Origin of Particles and the
Universe, Nagoya University, Nagoya 464-8602, Japan}

\begin{abstract}
For a spin-zero meson, the exact trace identity relates the D term to the
tensor gravitational form factor $A_M(t)$ and the total scalar trace
$\Theta_M(t)$.  We study the pion and kaon D terms by combining a holographic
tensor form factor with a chiral-dispersive representation of the trace.  Its
normalization and slope at the origin are obtained from the forward Ward
identity and curved-space SU(3) chiral perturbation theory.  A two-channel
$\pi\pi/K\bar K$ Muskhelishvili--Omn\`es solution constructed from empirical
scattering amplitudes describes the continuation to spacelike momentum.
Available lattice-QCD results are shown for comparison.  In
the chiral limit, saturation of the trace by the normalized explicit
quark-mass response gives $D_\pi(0)=-1/3$, while the soft-pion theorem gives
$D_\pi(0)=-1$.  This comparison separates the explicit-mass contribution
from the remaining chiral scalar response.  At physical masses, SU(3)
breaking in the chiral matching distinguishes the pion and kaon forward
values, and coupled-channel rescattering governs their evolution away from
the origin.  The kaon D term is less negative than the pion result in the
low-$Q^2$ spacelike region considered, with the separation decreasing as $Q^2$
increases.  These form factors provide a low-energy reference for future
lattice studies of the tensor and scalar channels.
\end{abstract}


\maketitle

\section{Introduction}
\label{sec:introduction}

The energy-momentum tensor (EMT) is the conserved current associated with
space-time translations.  Its forward matrix element fixes the hadron
four-momentum, while non-forward matrix elements describe the response to an
external metric.  Their momentum dependence is encoded in gravitational form
factors (GFFs), which are related to moments of generalized parton
distributions and enter hard exclusive processes
\cite{Ji:1996nm,Diehl:2003ny,Belitsky:2005qn,Kumano:2017lhr}.  GFFs also underlie discussions
of hadronic mass and mechanical distributions
\cite{Pagels:1966,Polyakov:2002,Polyakov:2018,JiMass:1995,
YangMass:2018,LiuMass:2021,LorceSchweitzer:2025}.

Two form factors are needed for a spin-zero hadron.  We denote them by $A(t)$
and $D(t)$.  The normalization $A(0)=1$ follows from momentum conservation,
whereas the D term is governed by the scalar part of the EMT and has no
corresponding normalization condition.  Lattice QCD now provides both pion
form factors for $0\leq-t<2\,\GeV^2$ at
$m_\pi\simeq170\,\mathrm{MeV}$~\cite{Hackett:2023}.  Continuum
bound-state calculations, chiral models, light-cone methods and resonance
descriptions offer complementary results for the pion and kaon
\cite{Broniowski:2008,Freese:2019,Aliev:2020,Krutov:2022,Xing:2022,
Xu:2024,Sultan:2024,Broniowski:2024,Yao:2025,Choi:2025}.
Recent studies have also considered scalar and dilaton descriptions of the
pion D term~\cite{XingSigma:2025,StegemanZwicky:2026,ZwickyDilaton:2026}.

The tensor channel has a simple realization in holographic QCD.  In the
hard-wall model, flavor gauge fields and a bifundamental scalar propagate on a
finite interval of AdS$_5$~\cite{Maldacena:1997re,Erlich:2005qh}.  A
transverse-traceless (TT) metric perturbation sources the spin-two projection
of the EMT, and its analytic bulk-to-boundary propagator turns the tensor form
factor into an overlap with normalizable hadron modes
\cite{Abidin:2008ku,Abidin:2008hn}.  Applications to mesons and baryons have
shown that this overlap gives a useful account of flavor breaking
\cite{Liu:2025,Liu:2026baryon}; related scalar and tensor currents have been
studied in light-front and top-down holographic approaches
\cite{Li:2024,Fujii:2024,WangLFHQCD:2024,Deng:2026}.

The TT projection provides the tensor form factor $A_M(t)$.  The D term also
involves the scalar form factor
$\Theta_M(t)=\langle M|T^\mu_{\ \mu}|M\rangle$ through the trace identity.
For the pion, the explicit quark-mass contribution and the total trace have
different slopes in the chiral limit.  As a reference, we assign the total
trace the normalized momentum dependence of the quark-mass matrix element.
This mass-shape saturation gives $D_\pi(0)=-1/3$ under the assumptions stated
in section~\ref{subsec:nogo}, while current algebra gives $D_\pi(0)=-1$ for
the total matrix element
\cite{Raman:1971,LeutwylerShifman:1989,Xing:2022,ZwickyDilaton:2026}.
The difference between these values is associated with the chiral response
beyond the chosen quark-mass reference shape.

The low-energy scalar response can be described by chiral perturbation theory
(ChPT) in a background metric.  The curvature couplings
$L_{11}$, $L_{12}$ and $L_{13}$ contribute to the flat-space EMT after metric
variation and generate the finite-mass corrections to the soft-pion result
\cite{GasserLeutwyler:1985,Donoghue:1991,KubisMeissner:1999,Hudson:2017}.
The same couplings enter the kaon form factors and encode finite-mass
SU(3)-breaking effects.  Away from the origin, the
scalar channel is influenced by $\pi\pi\leftrightarrow K\bar K$
rescattering.  We describe this momentum dependence with a coupled-channel
dispersive representation.  The required amplitudes are constrained by Roy and
Roy--Steiner analyses~\cite{GarciaMartin:2011,PelaezRodas:2022}, and dispersive
constructions of meson GFFs have recently been developed along these
lines~\cite{CaoExpanded:2025}.  Related dispersive methods have also been applied to
nucleon GFFs~\cite{CaoNature:2025}.

The dispersive analysis of Ref.~\cite{CaoExpanded:2025} treats the tensor and
trace channels together.  Here we follow a complementary route: the tensor
form factor is taken from the hard-wall model, while the dispersive equation
is applied to the total trace.  The two quantities are combined through the
exact trace identity.  This organization also permits separate comparisons
with future lattice results: $A_K$ probes the holographic tensor overlap,
whereas the pair $(A_K,D_K)$ determines the kaon trace.  The calculation is
thus formulated as a matched representation in a common EMT convention.

In this paper, we follow our previous holographic study of the pion and kaon
tensor form factors and include the scalar trace through a two-channel
Omn\`es representation.  Its subtraction constants are obtained from the
forward Ward identity and the correlated SU(3) ChPT values of $D_\pi(0)$ and
$D_K(0)$.  Available lattice form factors are shown for comparison and are
not used in the matching.  In the present calculation, holographic QCD is
used for the tensor channel, while ChPT and the dispersive representation are
employed for the low-energy scalar channel.  The forward D terms are fixed by
the ChPT matching, and their momentum dependence follows after the tensor and
trace channels have been combined.

Section~\ref{sec:formalism} introduces the EMT convention and the hard-wall
tensor form factor.  The explicit-mass trace, its chiral-limit relation and
the scalar sector of the hard-wall model are discussed in
section~\ref{sec:scalarproblem}.
Section~\ref{sec:trace} contains the ChPT matching, the two-channel Omn\`es
solution and the pion and kaon form factors.  We summarize the results in
section~\ref{sec:conclusions}.

\section{Holographic tensor form factor}
\label{sec:formalism}

We evaluate the tensor form factor in the hard-wall realization of the flavor
sector, using the parameters fixed in our earlier meson analysis.  The
normalizable pion and kaon modes are obtained from the flavor action.  Their coupling
to a transverse-traceless metric perturbation gives the three-point function
and the overlap representation for $A_M(t)$.

\subsection{EMT decomposition and trace identity}
\label{subsec:identity}

For the total, symmetric, and conserved EMT, Lorentz covariance leaves two
independent form factors between spin-zero states.  Let $p$ and $p'$ be the
on-shell momenta of a pseudoscalar meson $M$, and define
$P=(p+p')/2$, $\Delta=p'-p$ and $t=\Delta^2=-Q^2$.  We use the convention
\begin{equation}
 \begin{aligned}
 \langle M(p')|T^{\mu\nu}(0)|M(p)\rangle
 &=2P^\mu P^\nu A_M(t)\\
 &\quad+\frac12\left(\Delta^\mu\Delta^\nu
 -g^{\mu\nu}\Delta^2\right)D_M(t).
 \end{aligned}
 \label{eq:emt}
\end{equation}
The on-shell relation $P\cdot\Delta=0$ makes both structures transverse with respect to
$\Delta_\mu$.  No additional $g^{\mu\nu}$ form factor is present for the
conserved total EMT.  At zero momentum transfer, the corresponding charge is
the four-momentum operator.  With the state normalization implicit in
Eq.~\eqref{eq:emt}, translation invariance gives
\begin{equation}
 A_M(0)=1.
 \label{eq:A0}
\end{equation}
The trace probes a different linear combination of the same two form factors.
Using $P^2=m_M^2-t/4$ in Eq.~\eqref{eq:emt}, one finds the exact kinematic
identity
\begin{equation}
 \begin{aligned}
 \Theta_M(t)&\equiv
 \langle M(p')|T^\mu_{\ \mu}(0)|M(p)\rangle\\
 &=\left(2m_M^2-\frac{t}{2}\right)A_M(t)
 -\frac{3t}{2}D_M(t).
 \end{aligned}
 \label{eq:traceidentity}
\end{equation}
Its forward limit is fixed independently of hadron structure,
\begin{equation}
 \Theta_M(0)=2m_M^2.
 \label{eq:theta0}
\end{equation}
Solving Eq.~\eqref{eq:traceidentity} for the D term gives, in the spacelike
region,
\begin{equation}
 \begin{aligned}
 D_M(-Q^2)=-\frac{2}{3Q^2}\bigg[&
 \left(2m_M^2+\frac{Q^2}{2}\right)A_M(-Q^2)\\
 &-\Theta_M(-Q^2)\bigg].
 \end{aligned}
 \label{eq:Dmaster}
\end{equation}
The forward normalization in Eq.~\eqref{eq:theta0} removes the apparent
$0/0$ limit of Eq.~\eqref{eq:Dmaster}.  Expanding about the origin,
\begin{equation}
 \begin{aligned}
 A_M(-Q^2)&=1-a_MQ^2+\Order(Q^4),\\
 \Theta_M(-Q^2)&=2m_M^2-\dot\Theta_M(0)Q^2
 +\Order(Q^4),
 \end{aligned}
 \label{eq:expansions}
\end{equation}
where $a_M$ is the spacelike tensor slope and a dot denotes differentiation
with respect to timelike $t$ at the origin.  Substituting these expansions in
Eq.~\eqref{eq:Dmaster}, or equivalently applying l'H\^opital's rule, gives
\begin{equation}
 \begin{aligned}
 \dot\Theta_M(0)&=-\frac32D_M(0)-\frac12+2m_M^2a_M,\\
 D_M(0)&=-\frac23\left[\dot\Theta_M(0)+\frac12
 -2m_M^2a_M\right].
 \end{aligned}
 \label{eq:slopeidentity}
\end{equation}
The holographic TT overlap determines $a_M$, whereas ChPT and the dispersive
representation fix $\dot\Theta_M(0)$.

\subsection{Pseudoscalar modes}
\label{subsec:hardwall}

We use the standard $SU(3)_L\times SU(3)_R$ hard-wall flavor sector for the
meson wave functions.  The gauge fields $L_M$ and
$R_M$ are dual to the left- and right-handed flavor currents, while the
bifundamental field $X$ is dual to $\bar q_Rq_L$.  We set the AdS radius to
unity and restrict the fifth coordinate to $0<z\leq z_0$, with metric
\begin{equation}
 \dd s^2=\frac{1}{z^2}
 \left(\eta_{\mu\nu}\dd x^\mu\dd x^\nu-\dd z^2\right).
 \label{eq:adsmetric}
\end{equation}
The ultraviolet boundary is approached as $z\to0$, and the wall at $z=z_0$
sets the confinement scale.  The flavor action is~\cite{Erlich:2005qh}
\begin{equation}
 \begin{aligned}
 S_{\rm HW}=\int\dd^5x\sqrt{g}\,\tr\bigg[&
 (D_MX)^\dagger D^MX+3X^\dagger X\\
 &-\frac{F_{L,MN}F_L^{MN}+F_{R,MN}F_R^{MN}}
 {4g_5^2}\bigg].
 \end{aligned}
 \label{eq:hwaction}
\end{equation}
Here $D_MX=\partial_MX-iL_MX+iXR_M$, and $F_{L,R}$ are the corresponding
field strengths.  With the sign convention of Eq.~\eqref{eq:hwaction}, the
AdS/CFT relation
$m_5^2=\Delta(\Delta-4)$ gives $m_X^2=-3$ for the dimension-three operator
$\bar q_Rq_L$.  The remaining normalization is fixed by matching the
ultraviolet vector-current correlator,
\begin{equation}
 g_5^2=\frac{12\pi^2}{N_c}.
 \label{eq:g5}
\end{equation}
In the vacuum, only the scalar field is nonzero.  Its diagonal solution is
\begin{equation}
 \begin{aligned}
 X_0(z)&=\frac12\,\mathrm{diag}
 [v_q(z),v_q(z),v_s(z)],\\
 v_f(z)&=m_f\zeta z+\frac{\sigma_f}{\zeta}z^3,\\
 \zeta&=\frac{\sqrt{N_c}}{2\pi}.
 \end{aligned}
 \label{eq:vev}
\end{equation}
The coefficients of $z$ and $z^3$ are the quark-mass source and condensate,
respectively.  We retain the parameter set of the earlier hard-wall kaon
calculation,
\begin{equation}
 \begin{aligned}
 z_0^{-1}&=0.3225\,\GeV,\\
 m_q&=0.00831\,\GeV,\qquad m_s=0.1885\,\GeV,\\
 \sigma_q&=\sigma_s=(0.2137\,\GeV)^3.
 \end{aligned}
 \label{eq:hwinputs}
\end{equation}
These quantities were determined from the flat-space meson spectrum and decay
constants.  They are not readjusted in the present EMT calculation, so flavor
breaking in the tensor form factor is inherited from the same background that
describes the pion and kaon sectors.

For the pseudoscalar modes, we write
$X=X_0\exp(2i\pi^at^a)$ and introduce the axial combination
$A_M=(L_M-R_M)/2$.  Its four-dimensional component is decomposed as
$A_\mu=A_\mu^\perp+\partial_\mu\phi$.  The longitudinal field $\phi$ and the
phase $\pi$ mix in the symmetry-broken background; neither field alone is a
physical pseudoscalar mode.  Expanding Eq.~\eqref{eq:hwaction} to quadratic
order, choosing $A_z=0$, and inserting a four-dimensional mode of mass $m_M$
gives
\begin{align}
 \partial_z\left(\frac{1}{z}\partial_z\phi_M\right)
 +\frac{g_5^2[M_A^M(z)]^2}{z^3}\left(\pi_M-\phi_M\right)&=0,
 \label{eq:mode1}\\
 m_M^2\partial_z\phi_M
 -\frac{g_5^2[M_A^M(z)]^2}{z^2}\partial_z\pi_M&=0.
 \label{eq:mode2}
\end{align}
All flavor dependence in these equations is carried by the axial background
mass $M_A^M(z)$, which is fixed by the anticommutator of the axial generator with
$X_0$.  For the pion and kaon channels,
\begin{equation}
 [M_A^\pi(z)]^2=v_q^2(z),\qquad
 [M_A^K(z)]^2=\frac14[v_q(z)+v_s(z)]^2.
 \label{eq:axialmass}
\end{equation}
At the ultraviolet boundary, a normalizable solution contains no external
pseudoscalar or longitudinal-axial source.  Its overall amplitude remains
arbitrary.  We integrate the regular solution toward the infrared boundary
and impose
$\partial_z\phi_M(z_0)=0$; Eq.~\eqref{eq:mode2} then gives the corresponding
condition on $\partial_z\pi_M$.  Nontrivial solutions exist only for discrete
values of $m_M$, yielding
\begin{equation}
 m_{\pi,\rm HW}=0.13977\,\GeV,\qquad
 m_{K,\rm HW}=0.49579\,\GeV.
 \label{eq:hwmasses}
\end{equation}
The bulk wave functions use the eigenvalues in Eq.~\eqref{eq:hwmasses}, while
the four-dimensional kinematics are evaluated with the physical pion and kaon
masses.

Substituting a normalizable mode into the quadratic action fixes its
normalization.  After rescaling the four-dimensional meson field to unit
residue, the radial density is
\begin{equation}
 \begin{aligned}
 {\cal W}_M(z)=\frac{1}{{\cal N}_M}\bigg[&
 \frac{(\partial_z\phi_M)^2}{g_5^2z}
 +\frac{[M_A^M(z)]^2}{z^3}\\
 &\times(\pi_M-\phi_M)^2\bigg],\\
 &\int_0^{z_0}\dd z\,{\cal W}_M(z)=1.
 \end{aligned}
 \label{eq:weight}
\end{equation}
The two terms in ${\cal W}_M$ come from the axial gauge kinetic term and the
covariant derivative of $X$.  The same density appears in the TT variation of
the action and in the normalization of the meson pole.

\subsection{Transverse-traceless response}
\label{subsec:tt}

The boundary metric sources the total EMT.  We introduce a
transverse-traceless perturbation to isolate its spin-two projection,
\begin{equation}
 \begin{aligned}
 g_{\mu\nu}(x,z)&=\frac{1}{z^2}
 [\eta_{\mu\nu}+h_{\mu\nu}^{\rm TT}(x,z)],\\
 \partial^\mu h_{\mu\nu}^{\rm TT}&=0,
 \qquad h^{\rm TT\,\mu}_{\ \ \ \mu}=0.
 \end{aligned}
 \label{eq:ttsource}
\end{equation}
Contracting Eq.~\eqref{eq:emt} with this source gives
\begin{equation}
 h^{\rm TT}_{\mu\nu}
 \langle M(p')|T^{\mu\nu}|M(p)\rangle
 =2h^{\rm TT}_{\mu\nu}P^\mu P^\nu A_M(t).
 \label{eq:TTprojection}
\end{equation}
Transversality removes the $\Delta^\mu\Delta^\nu$ term, while tracelessness
removes the metric term.  The TT three-point function therefore determines
$A_M(t)$.  The complementary traceful projection enters through the scalar
channel discussed below.

The bulk profile of this source follows from the Einstein--Hilbert action in
the fixed AdS background,
\begin{equation}
 S_{\rm grav}=\frac{1}{2\kappa_5^2}\int\dd^5x\sqrt{g}\,(R+12)
 +S_{\rm GH}+S_{\rm IR}.
 \label{eq:gravAction}
\end{equation}
The cosmological term corresponds to a unit AdS radius.  The
Gibbons--Hawking and infrared terms complete the variational problem but do
not alter the bulk TT equation.  In the gauge of Eq.~\eqref{eq:ttsource}, the
quadratic action for a Euclidean Fourier mode is, up to the common
four-dimensional boundary volume,
\begin{equation}
 S_{\rm TT}^{(2)}=\frac{1}{8\kappa_5^2}
 \int_0^{z_0}\frac{\dd z}{z^3}
 \left[\partial_zh_{\mu\nu}^{\rm TT}\partial_zh^{{\rm TT}\,\mu\nu}
 +Q^2h_{\mu\nu}^{\rm TT}h^{{\rm TT}\,\mu\nu}\right].
 \label{eq:TTquadratic}
\end{equation}
Writing
$h_{\mu\nu}^{\rm TT}(q,z)=h_{\mu\nu}^{(0)}(q)H(Q,z)$ and varying
Eq.~\eqref{eq:TTquadratic} gives
\begin{equation}
 \begin{aligned}
 \partial_z\left(\frac{1}{z^3}\partial_zH\right)
 -\frac{Q^2}{z^3}H&=0,\\
 H''-\frac{3}{z}H'-Q^2H&=0.
 \end{aligned}
 \label{eq:Heq}
\end{equation}
No bulk mass parameter appears in Eq.~\eqref{eq:Heq}.  The metric perturbation
is the massless five-dimensional graviton dual to the conserved EMT, whereas
the scalar field $X$ has $m_X^2=-3$.

The ultraviolet Dirichlet condition fixes the normalization of the external
metric.  At $z=z_0$, the boundary variation of
Eq.~\eqref{eq:TTquadratic} is proportional to
$z_0^{-3}\partial_zH\,\delta H$.  In the absence of an infrared-localized
tensor source, it vanishes under the Neumann condition
\begin{equation}
 H(Q,0)=1,\qquad \partial_zH(Q,z_0)=0.
 \label{eq:Hbc}
\end{equation}
The two independent Euclidean solutions are proportional to
$Q^2z^2K_2(Qz)$ and $Q^2z^2I_2(Qz)$.  Applying the two boundary conditions
selects the combination
\begin{equation}
 H(Q,z)=\frac{Q^2z^2}{2}\left[
 K_2(Qz)+\frac{K_1(Qz_0)}{I_1(Qz_0)}I_2(Qz)
 \right].
 \label{eq:Hsolution}
\end{equation}
Expanding the flavor action to first order in $h_{\mu\nu}^{\rm TT}$ and to
second order in a canonically normalized pseudoscalar mode gives, with the
momentum-conserving delta function suppressed,
\begin{equation}
 \begin{aligned}
 S_{\rm TT}^{(3)}={}&
 \int\frac{\dd^4p\,\dd^4p'}{(2\pi)^8}
 \Phi_M^*(p')\Phi_M(p)\\
 &\times h_{\mu\nu}^{(0)}(\Delta)P^\mu P^\nu
 \int_0^{z_0}\dd z\,H(Q,z){\cal W}_M(z).
 \end{aligned}
 \label{eq:cubicoverlap}
\end{equation}
The axial kinetic term and the scalar covariant derivative reproduce exactly
the two contributions to ${\cal W}_M$ in Eq.~\eqref{eq:weight}.  Comparing
Eq.~\eqref{eq:cubicoverlap} with the boundary coupling
$\tfrac12h_{\mu\nu}^{(0)}\langle T^{\mu\nu}\rangle$, using
Eq.~\eqref{eq:TTprojection}, gives
\begin{equation}
 A_M^{\rm hol}(-Q^2)=\int_0^{z_0}\dd z\,
 H(Q,z){\cal W}_M(z).
 \label{eq:Aoverlap}
\end{equation}
Since $H(0,z)=1$, the normalization of ${\cal W}_M$ ensures that
$A_M^{\rm hol}(0)=1$.  The long-wavelength expansion of the graviton kernel is
\begin{equation}
 H(Q,z)=1+Q^2\left(-\frac{z^2}{4}+\frac{z^4}{8z_0^2}\right)
 +\Order(Q^4),
 \label{eq:Hexpansion}
\end{equation}
which gives
\begin{equation}
 a_M=-\left.\frac{\dd A_M^{\rm hol}(-Q^2)}{\dd Q^2}\right|_{0}
 =\int_0^{z_0}\dd z\,{\cal W}_M(z)
 \left(\frac{z^2}{4}-\frac{z^4}{8z_0^2}\right).
 \label{eq:aslope}
\end{equation}
The slope $a_M$ is therefore a radial moment of the same density that
normalizes the meson state.  For the parameter set in
Eq.~\eqref{eq:hwinputs},
\begin{equation}
 a_\pi=0.56407\,\GeV^{-2},\qquad
 a_K=0.52043\,\GeV^{-2}.
 \label{eq:aslopes}
\end{equation}

A traceful metric perturbation belongs instead to the scalar sector, where it
mixes with fluctuations of $X$ and with boundary terms.

\section{The scalar trace}
\label{sec:scalarproblem}

\subsection{Explicit-mass trace and the chiral limit}
\label{subsec:nogo}

The bifundamental field $X$ sources $\bar q q$ and therefore probes the
explicit quark-mass contribution to the trace.  We define
\begin{align}
 \Gamma_{m,M}(t)&=\left\langle M(p')\left|
 \sum_q m_q\bar q q\right|M(p)\right\rangle,
 \label{eq:GammaMass}\\
 \Gamma_{m,M}(0)&=\sum_q m_q\frac{\partial m_M^2}{\partial m_q},
 \label{eq:GammaFH}
\end{align}
where the forward limit follows from the Feynman--Hellmann theorem.  Its
normalized form factor is
\begin{equation}
 F_{m,M}(t)=\frac{\Gamma_{m,M}(t)}{\Gamma_{m,M}(0)},
 \qquad F_{m,M}(0)=1.
\end{equation}
Since $\Gamma_{m,M}(0)$ is not, in general, equal to the total-trace
normalization $2m_M^2$, $F_{m,M}$ and the full trace form factor describe
different matrix elements.  A useful reference for their momentum dependence
is the saturation ansatz
\begin{equation}
 \Theta_M^{\rm sat}(-Q^2)=2m_M^2F_{m,M}(-Q^2).
 \label{eq:mintrace}
\end{equation}
The rescaling satisfies the forward Ward identity without changing the
operator content of $\Gamma_{m,M}$.  For a finite mass radius,
$\dot\Theta_\pi^{\rm sat}(0)=\Order(m_\pi^2)$, and
Eq.~\eqref{eq:slopeidentity} reduces in the chiral limit to
\begin{equation}
 0=-\frac32D_\pi^{\rm sat}(0)-\frac12,
 \qquad
 D_\pi^{\rm sat}(0)=-\frac13.
 \label{eq:minus-third}
\end{equation}

The limit is taken with
\begin{equation}
 A_\pi(0)=1,\qquad
 \Theta_\pi(0)=2m_\pi^2,\qquad
 \dot\Theta_\pi(0)=\Order(m_\pi^2),
 \label{eq:minimalAssumptions}
\end{equation}
and with a tensor slope that remains finite as $m_\pi\to0$.  The value
$-1/3$ belongs to the saturation ansatz, independently of the detailed
holographic wave function.  The same chiral-limit value appears in other
symmetry-preserving pseudoscalar calculations
\cite{Xing:2022,Sultan:2024,XingSigma:2025}.

For the total QCD trace, the chiral Ward identities give
the limit~\cite{LeutwylerShifman:1989,Donoghue:1991,Hudson:2017,ZwickyDilaton:2026}
\begin{equation}
 D_\pi(0)\longrightarrow-1,\qquad
 \dot\Theta_\pi(0)\longrightarrow1
 \quad(m_\pi\to0).
 \label{eq:softpion}
\end{equation}
The comparison isolates the corresponding change in the scalar trace slope,
while the tensor normalization remains the same.

The non-normalizable mode of $X$ describes $\Gamma_{m,M}$, whereas the total
trace also contains the chiral response in Eq.~\eqref{eq:softpion} and the
gluonic anomaly.  In a holographic description, these contributions reside in
the coupled scalar metric--$X$ sector.

\subsection{The holographic scalar sector}
\label{subsec:sameaction}

A holographic description of the trace involves a dynamical
Einstein--scalar sector, with an action of the form
\begin{equation}
 \begin{aligned}
 S_{\rm dyn}={}&\int\dd^5x\sqrt{-g}\bigg[
 \frac{R_5}{2\kappa_5^2}
 -\frac{{\cal N}_X}{2}G_{IJ}\partial_M\phi^I\partial^M\phi^J\\
 &\hspace{24mm}-{\cal N}_X V(\phi)\bigg]
 +S_{\rm GH}+S_{\rm ct}+S_{\rm IR},
 \end{aligned}
 \label{eq:EinsteinScalarAction}
\end{equation}
with $\phi^I=(v_q,v_s)$.  Its fluctuation equations depend on
$\kappa_5^2{\cal N}_X$, the potential Hessian $V_{;IJ}$, and the second
variation of the infrared action.  These quantities constitute additional
scalar-sector inputs beyond the meson fit used for the TT sector.

Finite gravitational counterterms carry further low-energy information
\cite{Bianchi:2001kw}.  One allowed term is
\begin{equation}
 \Delta W_{\rm fin}=c_{13}\int\dd^4x\sqrt{-g}\,R[g]
 \left\langle\chi U^\dagger+U\chi^\dagger\right\rangle .
 \label{eq:finiteL13Audit}
\end{equation}
Although Eq.~\eqref{eq:finiteL13Audit} vanishes in flat space, its metric
variation contributes to the EMT and maps onto the chiral coupling $L_{13}$.
In the present analysis, the holographic background and TT sector are retained,
while the low-energy trace is described by ChPT and coupled-channel unitarity.

\section{Chiral matching and dispersive continuation}
\label{sec:trace}

\subsection{Total QCD trace}

We work throughout with the renormalized QCD Hilbert EMT.  Its trace contains
the quark-mass operators and the gluonic
trace anomaly~\cite{CollinsTrace:1977,NielsenTrace:1977},
\begin{equation}
 T^\mu_{\ \mu}=\sum_q(1+\gamma_m)m_q\bar qq
 +\frac{\beta(g)}{2g}G^a_{\mu\nu}G^{a\mu\nu}.
 \label{eq:anomaly}
\end{equation}
Recent studies have examined the pion trace anomaly and the gluon D form
factor in QCD factorization and in scalar-dilaton descriptions
\cite{CorianoTrace:2026,CorianoDilaton:2026,StegemanZwickyGluon:2026}.
Since the separation of the trace into quark-mass and finite-improvement form
factors depends on the effective operator basis, the dispersion relation is
written for the total scalar-isoscalar trace.  The two-channel spectral function describes its
low-energy part; unresolved higher-mass scalar and gluonic strength is absorbed
into the continuation above the empirical region.

Denoting the ChPT-matched dispersive trace by
$\Theta_M^{\rm ChPT+MO}$, we combine the tensor and trace inputs as
\begin{equation}
 \begin{aligned}
 A_M(t)&\equiv A_M^{\rm hol}(t),\\
 \Theta_M(t)&\equiv\Theta_M^{\rm ChPT+MO}(t),\\
 D_M(t)&\equiv
 D_M[A_M^{\rm hol},\Theta_M^{\rm ChPT+MO}](t).
 \end{aligned}
 \label{eq:matchingprescription}
\end{equation}
The D term then follows from Eq.~\eqref{eq:Dmaster}.  This defines a matched
spin-zero EMT representation in the chosen Hilbert-EMT convention, with the
tensor and trace form factors supplied by the holographic and
chiral-dispersive descriptions, respectively.

\subsection{Forward matching in curved-space ChPT}
\label{subsec:chpt}

At $\Order(p^4)$, the mesonic chiral Lagrangian in a background metric
contains~\cite{Donoghue:1991}
\begin{align}
 \mathcal L_{4}^{R}={}&L_{11}R\,
 \langle D_\mu U D^\mu U^\dagger\rangle
 +L_{12}R_{\mu\nu}\langle D^\mu U D^\nu U^\dagger\rangle \nonumber\\
 &+L_{13}R\,\langle\chi U^\dagger+U\chi^\dagger\rangle+\cdots .
 \label{eq:curvedchpt}
\end{align}
Although the curvature tensors vanish in flat space, their metric variations
contribute to the EMT matrix element.  We express the results of
Donoghue and Leutwyler for the total QCD Hilbert EMT in the convention of
Eq.~\eqref{eq:emt}.

At one loop, the physical forward D terms read
\begin{align}
 D_\pi(0)={}&-1+\frac{16m_\pi^2}{F_\pi^2}L_D^r
 +\frac{m_\pi^2}{F_\pi^2}I_\pi
 -\frac{m_\pi^2}{3F_\pi^2}I_\eta+\Order(p^4),
 \label{eq:DpiChPT}\\
 D_K(0)={}&-1+\frac{16m_K^2}{F_\pi^2}L_D^r
 +\frac{2m_K^2}{3F_\pi^2}I_\eta+\Order(p^4),
 \label{eq:DKChPT}\\
 L_D^r={}&L_{11}^r(\mu)-L_{13}^r(\mu),\\
 I_P={}&\frac{1}{48\pi^2}
 \left(\log\frac{\mu^2}{m_P^2}-1\right).
 \label{eq:chptloop}
\end{align}
The scale dependence of $L_D^r$ cancels that of the logarithms at this order.
At $\mu=1\,\GeV$, we take the reference input from the phenomenological
estimates of the curvature couplings
\cite{Donoghue:1991,CaoExpanded:2025},
\begin{equation}
 L_D^r=0.50\times10^{-3},\qquad F_\pi=92.1\,\mathrm{MeV},
\label{eq:LDinput}
\end{equation}
with $m_\pi=0.13957039\,\GeV$, $m_K=0.493677\,\GeV$, and
$m_\eta=0.547862\,\GeV$.  A common $L_D^r$ for the pion and kaon gives
\begin{equation}
 D_\pi(0)=-0.968,\qquad D_K(0)=-0.762.
 \label{eq:D0values}
\end{equation}
The corresponding trace slopes follow from Eq.~\eqref{eq:slopeidentity}; the
pion result approaches $-1$ as $m_\pi\to0$.

\subsection{Two-channel Omn\`es problem}
\label{subsec:mo}

For the right-hand cut, we use the CFD $\pi\pi$ phase of
Ref.~\cite{GarciaMartin:2011} and the dispersively constrained modulus and
phase of $g_0^0:\pi\pi\to K\bar K$ from Ref.~\cite{PelaezRodas:2022}.  The
published CFD polynomials are evaluated directly in the physical region.  No
closed parametrization or numerical table is available for the subthreshold
modulus, so we digitize the published curve and interpolate it to the kaon
threshold.  No GFF data enter this step or the matching.  The central result
uses the CFD$_B$ solution.

Our partial-wave normalization is
\begin{equation}
 S=\bm 1+2i\sqrt{\bm\Sigma}\,\bm T\sqrt{\bm\Sigma},\qquad
 \bm\Sigma=\operatorname{diag}(\sigma_\pi,\sigma_K),
 \label{eq:Snorm}
\end{equation}
with $\sigma_i(s)=\sqrt{1-4m_i^2/s}\,\theta(s-4m_i^2)$.  Above the kaon
threshold, the three functions $\delta(s)$, $|g(s)|$, and $\psi(s)$
determine an algebraically unitary symmetric matrix within a two-channel
closure,
\begin{align}
 T_{11}&=\frac{\eta e^{2i\delta}-1}{2i\sigma_\pi},
 &T_{12}=T_{21}&=|g|e^{i\psi},\nonumber\\
 T_{22}&=\frac{\eta e^{2i(\psi-\delta)}-1}{2i\sigma_K},
 &\eta&=\sqrt{1-4\sigma_\pi\sigma_K|g|^2}.
 \label{eq:Tmatrix}
\end{align}
Within this closure, $T_{22}$ above threshold is determined by the same three
functions.  Below the $K\bar K$ threshold, only the pion channel is open; we
use
$T_{11}=(e^{2i\delta}-1)/(2i\sigma_\pi)$,
$T_{12}=|g|e^{i\delta}$, and
$T_{22}=i\sigma_\pi|g|^2$.  Unitarity fixes
$\operatorname{Im}T_{22}=\sigma_\pi|T_{12}|^2$; its real part drops out of
$\bm T^\dagger\bm\Sigma$ because $\sigma_K=0$.  The isospin-symmetric
calculation uses a common kaon threshold.  The charged and neutral thresholds
near the physical $K\bar K$ region are discussed in
Ref.~\cite{PelaezRodas:2022}.

In the isoscalar S-wave, the pion and kaon trace form factors are collected in
\begin{equation}
 \bm\Theta(t)=
 \begin{pmatrix}
 \Theta^\pi(t)\\[2mm]
 \dfrac{2}{\sqrt3}\Theta^K(t)
 \end{pmatrix}.
 \label{eq:tracevector}
\end{equation}
With the isoscalar partial-wave normalization of
Ref.~\cite{Hoferichter:2012}, the component relations are
\begin{align}
 \operatorname{Im}\Theta_\pi
 &=\sigma_\pi T_{11}^*\Theta_\pi
 +\frac{2}{\sqrt3}\sigma_K T_{12}^*\Theta_K,
 \nonumber\\
 \operatorname{Im}\Theta_K
 &=\frac{\sqrt3}{2}\sigma_\pi T_{21}^*\Theta_\pi
 +\sigma_KT_{22}^*\Theta_K.
 \label{eq:componentunitarity}
\end{align}
Equivalently, in vector form,
\begin{equation}
 \begin{aligned}
 \operatorname{Im}\bm\Theta(t)
 &=\bm T_0^0(t)^*\bm\Sigma(t)\bm\Theta(t),\\
 \bm\Sigma(t)&=
 \begin{pmatrix}
 \sigma_\pi(t)\theta(t-4m_\pi^2)&0\\
 0&\sigma_K(t)\theta(t-4m_K^2)
 \end{pmatrix}.
 \end{aligned}
 \label{eq:unitarity}
\end{equation}
The Omn\`es matrix obeys the same discontinuity and is normalized at the
origin,
\begin{equation}
 \operatorname{Im}\bm\Omega_+(s)
 =\bm T(s)^\dagger\bm\Sigma(s)\bm\Omega_+(s),
 \qquad \bm\Omega(0)=\bm 1.
 \label{eq:Omegaunitarity}
\end{equation}
Writing $\bm\Omega_+=\bm X+i\bm Y$ and defining
\begin{equation}
 \bm A=\bm T^\dagger\bm\Sigma=\bm A_R+i\bm A_I,
 \qquad
 \bm R=(\bm 1+\bm A_I)^{-1}\bm A_R,
 \label{eq:Rmatrix}
\end{equation}
gives
\begin{equation}
 \bm Y(s)=\bm R(s)\bm X(s).
 \label{eq:YR}
\end{equation}
The canonical solution satisfies the unsubtracted dispersion
relation~\cite{Muskhelishvili:1953,Omnes:1958hv,Moussallam:1999,Hoferichter:2012}
\begin{align}
 \bm X(s)&=\frac{1}{\pi}\operatorname{PV}
 \int_{4m_\pi^2}^{\infty}\frac{\bm Y(s')}{s'-s}\,\dd s',
 \label{eq:canonicalPV}\\
 \bm 1&=\frac{1}{\pi}\int_{4m_\pi^2}^{\infty}
 \frac{\bm Y(s')}{s'}\,\dd s'.
 \label{eq:canonicalnorm}
\end{align}
The asymptotic index and the normalization in
Eq.~\eqref{eq:canonicalnorm} select the canonical solution.

We solve Eqs.~\eqref{eq:YR}--\eqref{eq:canonicalnorm} with continuous,
piecewise-linear basis functions on a nonuniform collocation grid.
Principal-value integrals are evaluated over the support of each basis
function, with the canonical
normalization imposed in the same linear system.

Above the empirical region, we continue the two-channel amplitude to its
asymptotic form.  Starting at $s_0=(1.30\,\GeV)^2$, the phases approach
$2\pi$ smoothly and the transition modulus approaches zero,
\begin{equation}
 \begin{aligned}
 x(s)&=2\pi+[x(s_0)-2\pi]f_n(s),\\
 |g(s)|&=|g(s_0)|f_n(s),\\
 f_n(s)&=\frac{2}{1+(\sqrt{s}/\sqrt{s_0})^n}.
 \end{aligned}
 \label{eq:asymptoticguide}
\end{equation}
We set $n=3$ and truncate the numerical integral at
$\sqrt{s_{\rm max}}=8\,\GeV$.

For spacelike momentum transfer, the same dispersion relation gives
\begin{align}
 \bm\Omega(-Q^2)&=\frac1\pi\int_{4m_\pi^2}^{\infty}
 \frac{\bm Y(s')}{s'+Q^2}\,\dd s',
 \label{eq:OmegaSpacelike}\\
 \dot{\bm\Omega}(0)&=\frac1\pi\int_{4m_\pi^2}^{\infty}
 \frac{\bm Y(s')}{s'^2}\,\dd s'.
 \label{eq:OmegaSlope}
\end{align}
The off-diagonal entries encode $\pi\pi\leftrightarrow K\bar K$ rescattering.
Their contribution to the kaon trace describes a scalar source that first
couples to $\pi\pi$, followed by rescattering into $K\bar K$.

The trace vector is written as
\begin{equation}
 \bm\Theta(t)=\bm\Omega(t)[\bm a+t\bm b],
 \qquad
 \bm a=\begin{pmatrix}2m_\pi^2\\4m_K^2/\sqrt3\end{pmatrix}.
\label{eq:MOtrace}
\end{equation}
We use the standard two-channel MO representation for meson trace form
factors~\cite{CaoExpanded:2025}.  The forward trace Ward identity fixes
$\bm a$; at NLO, the source polynomial is linear, with slope
\begin{equation}
 \dot\Theta_M(0)=-\frac32D_M(0)-\frac12+2m_M^2a_M,
 \label{eq:ThetaSlopeFromD}
\end{equation}
where $D_M(0)$ comes from ChPT and $a_M$ from the holographic tensor form factor.
The resulting slopes are
\begin{equation}
 \dot\Theta_\pi(0)=0.974,\qquad \dot\Theta_K(0)=0.897.
 \label{eq:matchedTraceSlopes}
\end{equation}
In the normalization of Eq.~\eqref{eq:tracevector},
\begin{equation}
 \dot{\bm\Theta}(0)=
 \begin{pmatrix}\dot\Theta_\pi(0)\\2\dot\Theta_K(0)/\sqrt3\end{pmatrix}.
\end{equation}

Differentiating Eq.~\eqref{eq:MOtrace} at the origin gives
\begin{equation}
 \bm b=\dot{\bm\Theta}(0)-\dot{\bm\Omega}(0)\bm a.
 \label{eq:bsource}
\end{equation}
Thus the source slope is fixed by the forward trace and the Omn\`es slope.
With the inputs specified above,
\begin{equation}
 \bm b=\begin{pmatrix}0.71644\\0.42176\end{pmatrix}.
 \label{eq:centralmatching}
\end{equation}
The term $\dot{\bm\Omega}(0)\bm a$ subtracts the slope already generated by
rescattering.

\subsection{Pion and kaon form factors}
\label{subsec:external}

We obtain the pion and kaon D terms by combining
Eqs.~\eqref{eq:Dmaster}, \eqref{eq:Aoverlap} and \eqref{eq:MOtrace}.  The
holographic parameters are inherited from the meson fit.  The coupling
$L_D^r$ fixes the correlated forward values, while the scattering amplitudes
determine the Omn\`es matrix.  The lattice points provide an external
comparison and are independent of the matching procedure.

At $Q^2=0.5\,\GeV^2$, we find
\begin{equation}
 \begin{aligned}
 D_\pi(-0.5\,\GeV^2)&=-0.570,\\
 D_K(-0.5\,\GeV^2)&=-0.428,\\
 \frac{D_K}{D_\pi}&=0.751.
 \end{aligned}
 \label{eq:headline}
\end{equation}

\begin{figure*}[t]
\centering
\includegraphics[width=\textwidth]{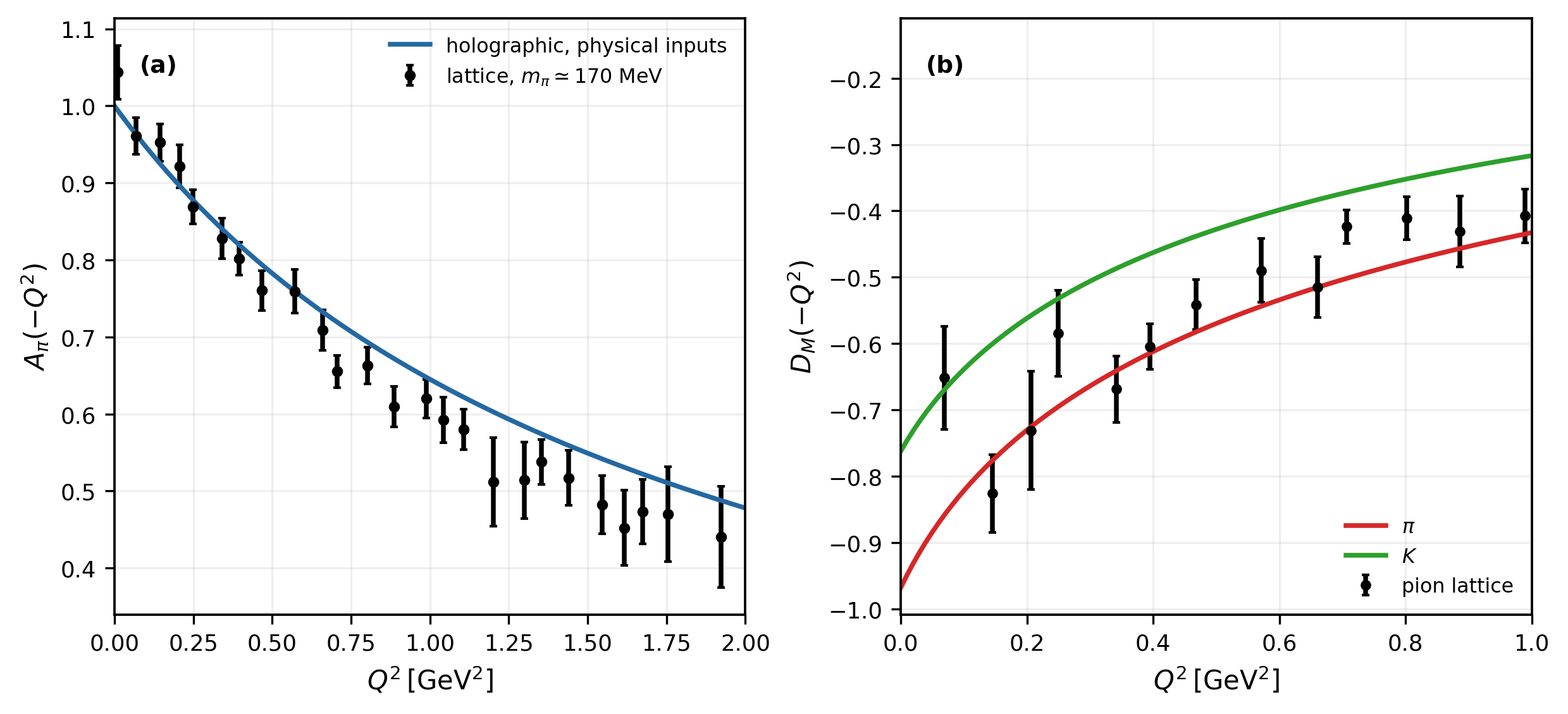}
\caption{Pion and kaon gravitational form factors.  Panel (a) compares the
holographic tensor form factor with the pion lattice-QCD results of
Ref.~\cite{Hackett:2023}.  Panel (b) shows the D terms obtained from the
two-channel trace matching.  The red and green lines denote the pion and kaon
results, respectively, and the black markers show the pion lattice-QCD data.
The lattice ensemble has $m_\pi\simeq170\,\mathrm{MeV}$, while the curves use
the physical inputs specified in Eq.~\eqref{eq:hwinputs}.}
\label{fig:main}
\end{figure*}

Figure~\ref{fig:main} shows the resulting momentum dependence.  The
holographic $A_\pi$ is compared with the lattice result of
Ref.~\cite{Hackett:2023}; the second panel displays the matched pion and kaon D
terms.  The lattice ensemble has $m_\pi\simeq170\,\mathrm{MeV}$, whereas the
curves use the physical inputs specified in Eq.~\eqref{eq:hwinputs}.

\section{Conclusion}
\label{sec:conclusions}

We have studied the pion and kaon EMT matrix elements by combining a
holographic calculation of the tensor form factor with a chiral-dispersive
representation of the total trace.  These quantities enter the D term through
the exact spin-zero trace identity.  In the holographic model, the
transverse-traceless graviton gives $A_M(t)$ as an overlap of the metric kernel
with the normalizable meson mode.  Curved-space ChPT and the coupled-channel
dispersion relation provide the low-energy scalar trace $\Theta_M(t)$.

The chiral limit of the pion offers a useful reference for separating the
scalar contributions.  The explicit quark-mass reference form factor and the
current-algebra result for the total EMT correspond to different scalar trace
slopes.  At physical masses, curved-space SU(3) ChPT supplies
the finite-mass corrections through the common combination
$L_D^r=L_{11}^r-L_{13}^r$ and yields distinct forward values for the pion and
kaon.

The finite-$t$ dependence of the trace was obtained from a two-channel
$\pi\pi/K\bar K$ Omn\`es matrix constructed from empirical scattering
amplitudes.  Its subtraction vector is fixed by the forward Ward identity and
the ChPT trace slopes.  Combining this representation with the holographic
tensor form factor determines the spacelike D terms.

For the reference input, the kaon D term remains less negative than the pion
D term over the spacelike interval shown in Fig.~\ref{fig:main}.  This
behavior follows from the correlated ChPT matching and the coupled-channel
trace.  The lattice pion points in the figure provide an independent
comparison with the resulting form factors.

The calculation defines a matched four-dimensional EMT representation.
Two-channel unitarity is implemented explicitly in the low-energy region,
while higher scalar channels and short-distance strength enter through the
continuation above the empirical scattering range.  This organization
preserves the connection between the tensor form factor and holographic meson
phenomenology
and incorporates the scalar response constrained by low-energy QCD.

Future lattice calculations of $A_K(t)$ and $D_K(t)$ would permit separate
comparisons with the tensor and trace parts of the analysis.  A three-channel
$\pi\pi/K\bar K/\eta\eta$ Omn\`es solution and a higher-order curved-space
ChPT determination of the subtraction constants would refine the scalar
trace.  A dynamical Einstein--scalar realization offers a complementary bulk
description.  Together, the pion and kaon form factors may help clarify how
the chiral scalar response evolves with the strange-quark mass.

\begin{acknowledgments}
This work was supported by the JSPS Research Fellowship for Young
Scientists (No. JP26KJ1338).
\end{acknowledgments}

\end{document}